\documentclass[12pt,preprint]{aastex}

\def\R23{$R_{23}$}

\def\O32{${\rm O}_{32}$}

\def\hst{{\it HST}}

\slugcomment{Submitted to ApJ}

\begin{document}

\title{Evidence for a Supernova Associated with the X-ray Flash
020903\footnote{Based on observations made with the NASA/ESA Hubble
Space Telescope, obtained at the Space Telescope Science Institute,
which is operated by the Association of Universities for Research in
Astronomy, Inc., under NASA contract NAS 5-26555. These observations
are associated with program GO-9405 (P.I. A. Fruchter).}}


\author{
D.~Bersier\altaffilmark{2},
A.~S.~Fruchter\altaffilmark{2},
L.-G.~Strolger\altaffilmark{2},
J.~Gorosabel\altaffilmark{2,3},
A.~Levan\altaffilmark{4},
I.~Burud\altaffilmark{2},
J.~E.~Rhoads\altaffilmark{2},
A.~C.~Becker\altaffilmark{5},
A.~Cassan\altaffilmark{6},
R.~Chornock\altaffilmark{7},
S.~Covino\altaffilmark{8},
R.~S.~de~Jong\altaffilmark{2},
D.~Dominis\altaffilmark{9,10},
A.~V.~Filippenko\altaffilmark{7},
J.~Hjorth\altaffilmark{11},
J.~Holmberg\altaffilmark{12},
D.~Malesani\altaffilmark{13},
B.~Mobasher\altaffilmark{2},
K.~A.~G.~Olsen\altaffilmark{14},
M.~Stefanon\altaffilmark{15},
J.~M.~Castro~Cer\'{o}n\altaffilmark{1,10}
J.~P.~U.~Fynbo\altaffilmark{10},
S.~T.~Holland\altaffilmark{16,17},
C.~Kouveliotou\altaffilmark{18},
H.~Pedersen\altaffilmark{10},
N.~R.~Tanvir\altaffilmark{19},
S.~E.~Woosley\altaffilmark{20}
}

\altaffiltext{2}{Space Telescope Science Institute, 3700 San Martin
Drive, Baltimore, MD 21218.}
\altaffiltext{3}{Instituto de Astrof\'{\i}sica de Andaluc\'{\i}a
(IAA-CSIC), P.O. Box 03004, E-18080 Granada, Spain.}
\altaffiltext{4}{Department of Physics and Astronomy, University of
Leicester, University Road, Leicester, LE1 7RH, UK.}
\altaffiltext{5}{Department of Astronomy, University of Washington,
Box 351580, Seattle, WA 98195.}
\altaffiltext{6}{Institut d'Astrophysique de Paris, 98bis Boulevard Arago,
75014 Paris, France.}
\altaffiltext{7}{Department of Astronomy, University of California,
Berkeley, CA 94720-3411.}
\altaffiltext{8}{INAF-Osservatorio Astronomico di Brera, Via Bianchi
46, I-23807 Merate (LC), Italy.}
\altaffiltext{9}{Universit\"at Potsdam, Am Neues Palais 10, D-14469
Potsdam, Germany.}
\altaffiltext{10}{Astrophysikalische Institut Potsdam, An der
Sternwarte 16, D-14482 Potsdam, Germany.}
\altaffiltext{11}{Niels Bohr Institute, University of Copenhagen,
Juliane Maries Vej 30, DK-2100, Copenhagen, Denmark.}
\altaffiltext{12}{Tuorla Observatory, V\"ais\"al\"antie 20, FI-21500,
Piikki\"o, Finland.}
\altaffiltext{13}{International School for Advanced Studies
(SISSA-ISAS), via Beirut 2-4, I-34014 Trieste, Italy.}
\altaffiltext{14}{Cerro Tololo Interamerican Observatory, National
Optical Astronomy Observatory, Casilla 603, La Serena, Chile.}
\altaffiltext{15}{INAF-Osservatorio Astronomico di Brera, Via Bianchi
46, I-23807 Merate (LC), Italy.}
\altaffiltext{16}{Goddard Space Flight Center, Code 660.1, Greenbelt, MD
20771-0003} 
\altaffiltext{17}{Universities Space Research Association.} 
\altaffiltext{18}{NASA Marshall Space Flight Center, NSSTC, SD-50, 320
Sparkman Drive, Huntsville, AL 35805} 
\altaffiltext{19}{Centre for Astrophysics Research, University of
Hertfordshire, College Lane, Hatfield AL10 9AB, UK} 
\altaffiltext{20}{Department of Astronomy and Astrophysics, University
of California, Santa Cruz, CA 95064} 

\begin{abstract}
We present ground-based and \emph{Hubble Space Telescope} optical
observations of the X-ray flash (XRF) 020903, covering 300 days.  The
afterglow showed a very rapid rise in the first day, followed by a
relatively slow decay in the next few days.  There was a clear bump in
the light curve after $\sim 25$ days, accompanied by a drastic change
in the spectral energy distribution. The light curve and the spectral
energy distribution are naturally interpreted as the emergence -- and
subsequent decay -- of a supernova (SN), similar to SN~1998bw.  At
peak luminosity, the SN is estimated to be $0.8 \pm 0.1$ mag fainter
than SN~1998bw.  This argues in favor of the existence of a supernova
associated with this X-ray flash.
A spectrum obtained 35 days after the burst shows emission lines from
the host galaxy. We use this spectrum to put an upper limit on the
oxygen abundance of the host at [O/H]$\leq -0.6$ dex.
We also discuss a possible trend between the softness of several
bursts and the early behavior of the optical afterglow, in the sense
that XRFs and X-ray rich GRBs seem to have a plateau phase or even a
rising light curve. This can be naturally explained in models where
XRFs are similar to GRBs but seen off the jet axis.
\end{abstract}

\keywords{gamma-ray bursts --- supernovae}

\section{Introduction}

X-ray flashes (XRFs) are bursts of high-energy light whose properties
are similar to those of gamma-ray bursts (GRBs): they are transient
events of short duration (typically a minute), distributed
isotropically on the sky, and have fluences, frequencies (number per
day), and spectra comparable to those of GRBs
\citep{heise2003,kippen2003}.  Their spectra are similar to those of
GRBs but their peak energies in the $\nu F_\nu$ spectrum, $E_{\mathrm
peak}$, are smaller than those of GRBs
\citep{heise2003,kippen2003,sakamoto2005}.  In various respects, XRFs
seem to be a low-energy extension of GRBs. In light of these
similarities, it is tempting to try to unify these classes of events
into a single framework (e.g., Lamb, Donaghy, \& Graziani, 2004, and
references therein).  Differences between XRFs and GRBs may be
intrinsic, caused for instance by different Lorentz factor
\citep{barraud2005} or by a high baryon load \citep{dermer1999}.  They
might be extrinsic; in particular the observed properties are
determined by the viewing angle
\citep{yamazaki2002,yamazaki2004,granot2005}.

The redshift of XRF~020903 ($z=0.251$; Soderberg et al., 2004; this
paper) clearly shows that XRFs are cosmic explosions of phenomenal
power, almost comparable to GRBs.  The isotropic energy of this XRF is
very large ($1.1\times 10^{49}$ erg) though not as large as that of
typical GRBs (up to $10^{54}$ erg; see Bloom, Frail \& Kulkarni, 2003).
The similarities and differences between XRFs and GRBs will hopefully
help us elucidate the nature of these events and perhaps find a common
model describing both phenomena. 

One obvious requirement for a unified model is that both types of
event (XRFS and GRBs) have the same kind of progenitor. For gamma-ray
bursts it is now clear that they are associated with core-collapse
supernovae (SN) of type Ib/c.  This GRB/SN connection has received
more and more support over the years.  Some theoretical models
(e.g. Woosley, 1993; see Zhang, Woosley, \& Heger, 2004, for recent
developments) predict such a connection for GRBs and XRFs, but it is
only with GRB~980425 that observations showed the first hint of this
association \citep{galama1998}. The bump on the light curve of
GRB~011121 and its rapid color evolution (e.g. Garnavich et al. 2003)
provided strong evidence in favor of this genetic link between GRBs
and SNe.  There have been other GRBs with ``bumps'' (``bump'' is here
defined as a rebrightening observed 15-20 days after the burst) in
their light curves that could be attributed to SNe but none were as
definitive as GRB~030329 \citep{stanek2003,hjorth03}; the spectrum of
this burst showed features that could only be explained with the
existence of an underlying Type Ib/c SN.  Later, \citet{malesani2004}
presented spectroscopic evidence for the existence of a SN associated
with GRB~031203.  In cases where a SN could have been detected, there
was indeed a bump on the light curve (See Zeh et al, 2004, and
references therein).

When it comes to XRFs however, the situation is less clear.
\citet{soderberg2005a} presented for several X-ray flashes.  Although
the conclusions of their study are limited by the fact that some
objects do not have a measured redshift, they do detect supernova
signatures in some cases. In contrast, they also put strong upper
limits on the brightness of any possible SN associated with some of
the events they studied.  \citet{fynbo2004} showed that XRF~030723
actually had a strong bump on its light curve that could be fit
reasonably well with the light curve of SN~1994I. At the time of the
bump the energy distribution was very red and was not consistent with
a power law.  Furthermore, \citet{levan2005a} report a non-detection of
supernova features in the light curves of XRF~011030 and XRF~020427.
All one can say from these works is that if XRFs are associated with
SNe, they cover a large range in peak magnitude. It is thus necessary
to observe XRFs thoroughly and determine the peak magnitude of any SN
associated with these events.

The burst XRF~020903 was detected by the \emph{High Energy Transient
Explorer 2} (HETE-2) on 3 September 2002 at 10:05:38 \citep{gcn1530}.
Optical observations started within a few hours after the burst.
Initially no afterglow was detected
\citep{gcn1531,gcn1533,gcn1535,gcn1537}.  Images taken 6 d apart on
the Cerro Tololo Inter-American Observatory (CTIO) Blanco 4~m
telescope did not show an obvious variable source \citep{gcn1557}.
The analysis we present here shows that the optical transient (OT)
varied very little between these two epochs. Using data obtained on 4
September and 10 September 2002, \citet{gcn1554} reported (26 days
after the burst) the discovery of an optical afterglow.  From optical
spectroscopy obtained on 28 September 2002, \citet{gcn1554} measured a
redshift $z = 0.25$; this was the first reported redshift for an XRF.
A fading radio source was also discovered (at the position
$\alpha_{2000} = 22^h 48^m 42\fs 339$, $\delta_{2000} = -20\arcdeg 46'
08\farcs 95$, Soderberg et al. 2002, as updated by Soderberg et
al. 2004), indicating an association with XRF~020903.  After about 30
days, there was no variability observed in the optical
\citep{gcn1563,gcn1631}, presumably because of the bright host galaxy.
Further {\it Hubble Space Telescope} (\hst) observations
\citep{gcn1761} showed that the source was actually still fading.

\citet{sakamoto2004} showed that the peak energy, $E_{\rm peak}$, of
XRF~020903 was very low, below $\sim 3$ keV.  From the redshift and
observed fluence, the equivalent isotropic energy, $E_{\rm iso}$, was
$1.1\times 10^{49}$ erg.  Furthermore, there was no detected emission
beyond 10 keV. XRF~020903 is thus one of the most extreme X-ray
flashes observed by HETE-2 (see Sakamoto et al. 2005, for a general
discussion of GRBs and XRFs found by HETE-2).

We obtained data on XRF~020903 using ground-based telescopes and the
{\it Hubble Space Telescope} (\hst); these observations are presented
in Sect.~\ref{sec_obs}. In Sect.~3, we discuss the light curve and the
spectral energy distribution; we also present a spectrum of the host
galaxy and determine an upper limit on its oxygen abundance. We
discuss our findings in Sect.~\ref{sec_concl}.

\section{Observations and Photometry\label{sec_obs}}



\subsection{Ground-Based Data}

We obtained $BRI$ data $\sim 0.6$~d after the burst with the
wide-field MOSAIC II camera on the CTIO Blanco 4~m telescope. We
secured another $R$-band observation with this telescope 5.7~d after
the burst.
 We also have $R$-band data, observed on 29 Sep 2002 with the Asiago
Faint Object Spectrograph and Camera (AFOSC) on the 1.82~m
``Copernicus'' telescope at Mt Ekar (Asiago, Italy), and we have $VRI$
data taken on 2 October 2002 with the 3.6~m Telescopio Nazionale
Galileo (TNG) at La Palma, where we used the DOLoRes camera
\citep{conconi2001}.  A preliminary analysis of these TNG and Asiago
data has been given by \citet{gcn1563}.  Between 10 and 14 October
2002, and then on 26 October, we obtained $BVRI$ data with the Danish
1.5~m telescope at La Silla Observatory \citep{gcn1631}.

\begin{figure}
\epsscale{.30}\plotone{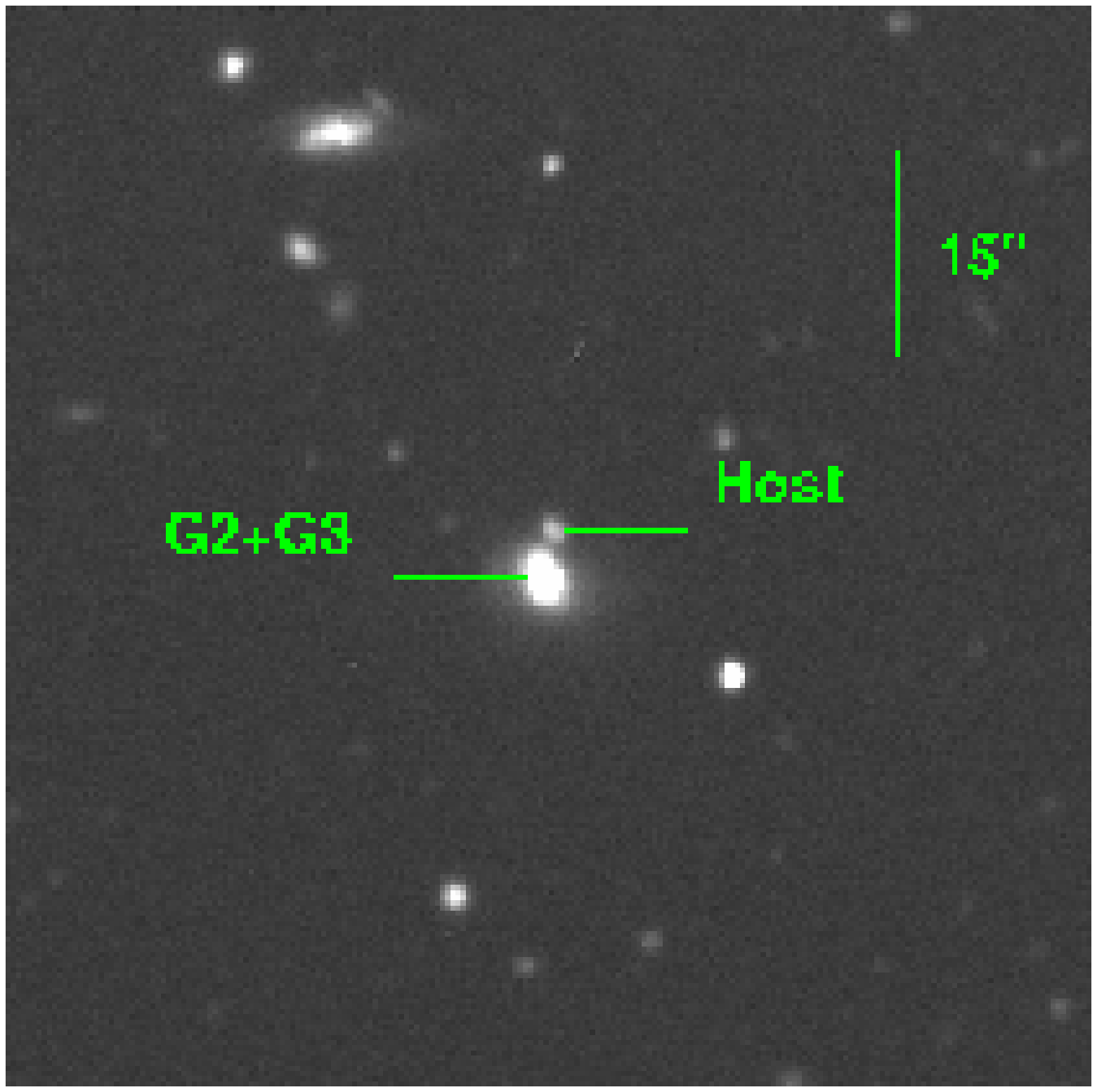}
\epsscale{.295}\plotone{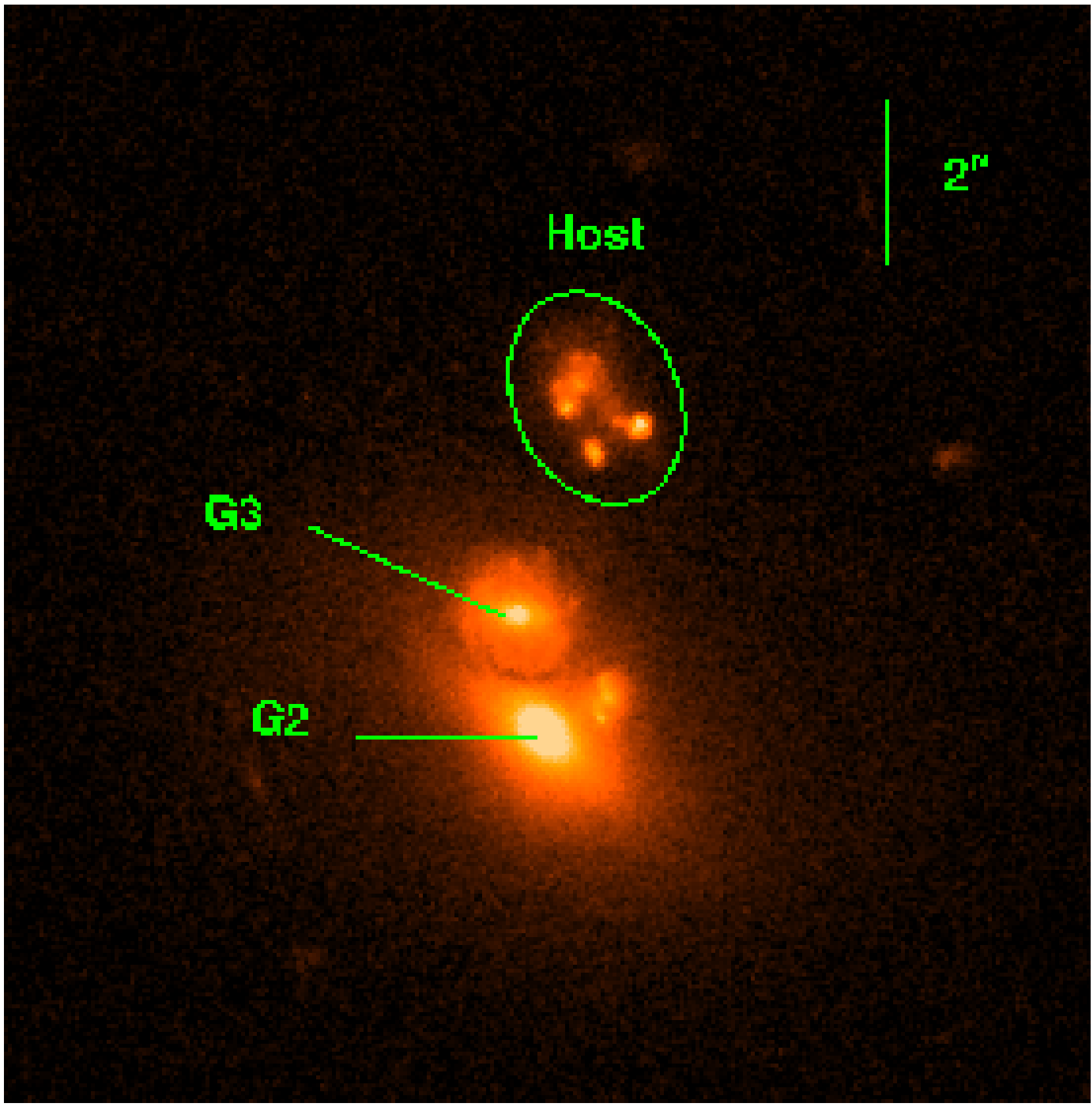}
\epsscale{.30}\plotone{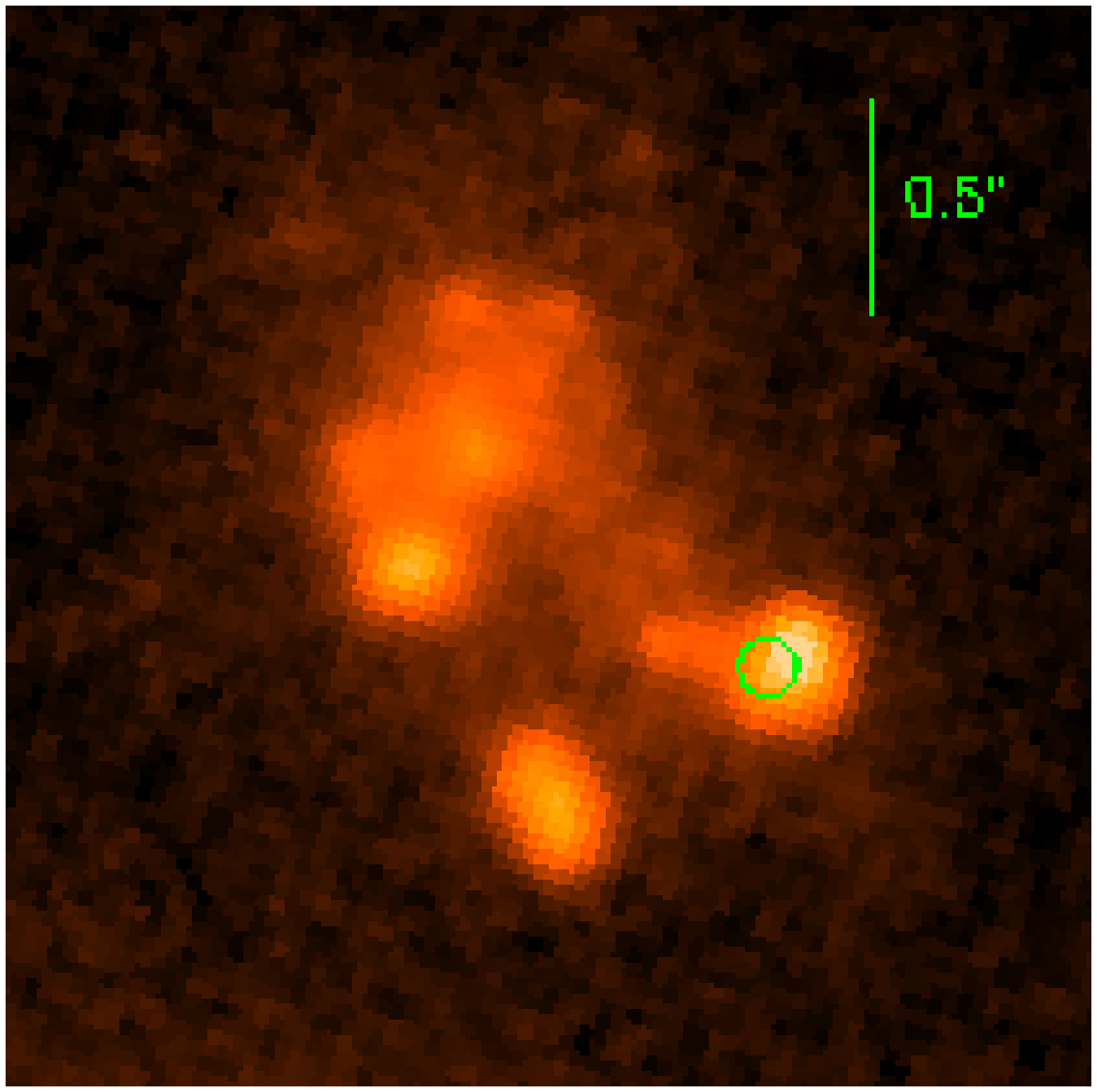}
\caption{{\it (Left panel)} CTIO $R$-band image of the area around
XRF~020903. North is up, east is to the left.  The host is the small
galaxy near the center of the image.  The galaxies G2 and G3 are not
resolved in our ground-based images.
{\it (Center panel)} \hst\ image showing the host (the cluster of
knots in the upper part of the image), and the nearby galaxies G2 and
G3.
{\it (Right panel)} \hst\ image showing in detail the complex structure
of the host. The $0\farcs 1$ circle is centered on the position of the
OT.
\label{fig_host}}
\end{figure}

Given that the OT is located in a fairly bright host (see
Fig.~\ref{fig_host}), the ground-based photometry is delicate. In
order to remove the contribution of the host galaxy, we used the image
convolution and subtraction methods of Alard (2000; see also Alard \&
Lupton 1998).  To fully exploit this method one needs a reference
image where the optical transient has faded well below the detection
limit.  We thus secured late-time images in August 2004 with the CTIO
4~m telescope (in $RI$), and with the Danish 1.5~m telescope (in
$BV$).  These images have been used as templates in the image
subtraction method.  We then performed aperture photometry on each
subtracted frame.  The absolute calibration was done using secondary
standards from the list of \citet{gcn1571}. Our photometry is
presented in Table~\ref{tbl_phot}.

\clearpage
\clearpage


\begin{table}
\begin{center}
\caption{Ground-based photometry of XRF 020903\tablenotemark{a}.\label{tbl_phot}}
\begin{tabular}{l c l c l r}
\tableline\tableline UT date & Time after burst (d) & %
Telescope\tablenotemark{b} & Filter & mag & $\sigma_{\rm mag}$  \\ 
\tableline 
2002 Sep 04.115                 &  0.695 & CTIO 4m    & B & 21.40   & 0.17 \\ 
2002 Oct 10.177                 & 36.757 & DK 1.5m    & B & 24.59   & 0.26 \\
2002 Oct 11.193                 & 37.772 & DK 1.5m    & B & 24.76   & 0.31 \\
2002 Oct 12.169                 & 38.749 & DK 1.5m    & B & 24.77   & 0.32 \\
2002 Oct 13.163                 & 39.743 & DK 1.5m    & B & $>$24.5 & - \\
2002 Oct 14.137                 & 40.716 & DK 1.5m    & B & $>$24.4 & - \\
\hline 
2002 Oct 02.953                 & 29.533 & TNG        & V & $>$22.8 & - \\
2002 Oct 10.192                 & 36.772 & DK 1.5m    & V & $>$23.4 & - \\
2002 Oct 11.7\tablenotemark{c}  & 38.279 & DK 1.5m    & V & 23.45   & 0.21 \\
2002 Oct 13.179                 & 39.758 & DK 1.5m    & V & $>$23.2 & - \\ 
2002 Oct 14.152                 & 40.731 & DK 1.5m    & V & $>$23.1 & - \\
\hline
2002 Sep 04.080                 &  0.660 & CTIO 4m    & R & 20.858  & 0.007 \\
2002 Sep 04.32\tablenotemark{d} &  0.900 & Palomar 5m & R & 19.52   & 0.22  \\
2002 Sep 09.130                 &  5.709 & CTIO 4m    & R & 21.129  & 0.010 \\
2002 Sep 10.30\tablenotemark{d} &  6.880 & Palomar 5m & R & 21.80   & 0.30  \\
2002 Sep 29.923                 & 26.503 & Asiago     & R & 21.729  & 0.089 \\
2002 Oct  2.967                 & 29.547 & TNG        & R & 21.552  & 0.044 \\
2002 Oct 10.204                 & 36.783 & DK 1.5m    & R & 22.351  & 0.051 \\
2002 Oct 11.216                 & 37.796 & DK 1.5m    & R & 22.279  & 0.052 \\
2002 Oct 12.193                 & 38.772 & DK 1.5m    & R & 22.450  & 0.055 \\
2002 Oct 13.187                 & 39.766 & DK 1.5m    & R & 22.522  & 0.054 \\
2002 Oct 14.110                 & 40.689 & DK 1.5m    & R & 22.498  & 0.065 \\
2002 Oct 26.062                 & 52.641 & DK 1.5m    & R & 23.032  & 0.087 \\
\hline
2002 Sep 04.107                 &  0.686 & CTIO 4m    & I & 20.425  & 0.015 \\
2002 Oct 02.977                 & 29.556 & TNG        & I & 20.916  & 0.062 \\ 
2002 Oct 10.172                 & 36.752 & DK 1.5m    & I & 21.430  & 0.064 \\
2002 Oct 12.201                 & 38.780 & DK 1.5m    & I & 21.442  & 0.060 \\
2002 Oct 13.071                 & 39.651 & DK 1.5m    & I & 21.495  & 0.057 \\
\tableline
\end{tabular}

\tablenotetext{a}{This is the photometry of the OT only, after
subtraction of the host galaxy.}

\tablenotetext{b}{The key to telescope identification is as follows:
CTIO 4m = Cerro Tololo 4~m, DK 1.5m = 1.5~m Danish telescope at La
Silla, TNG = 3.6~m Telescopio Nazionale Galileo at La Palma, Asiago =
1.8~m telescope at Mt Ekar, Palomar 5m = Mt Palomar 5~m Hale
telescope.}

\tablenotetext{c}{Average of two measurements on Oct 11 and Oct 12.}

\tablenotetext{d}{From \citet{soderberg2004}, after having subtracted
the contribution from the host galaxy (see text for details).  Note
that these data have been re-examined by \citet{soderberg2005a} where
they use image subtraction. According to their Figure~2, the
measurement at $\sim 6.9$~d is $\sim 21.8$ mag, the value we use
here.}

\tablecomments{The data in this table supersede the data presented by
\citet{gcn1557,gcn1631}, and \citet{gcn1761}.}

\end{center}
\end{table}

\subsection{{\it Hubble Space Telescope} Data}

We obtained data at three epochs (95, 232, and 301 days after the
burst) with \hst, using the Advanced Camera for Surveys (ACS) and the
F606W filter \citep{gcn1761}.  The images were reduced and drizzled in
a standard manner \citep{drizzle2002}.  The pixel scale of the final
images is $0\farcs 033$, $2/3$ of the native ACS scale ($0\farcs 05$).
The first- and second-epoch images were then aligned with the
third-epoch image. Subtracting the third image from the first epoch
gave a clear detection of the OT, and a very accurate position.

Given the brightness of the OT after $\sim 35$ days, it is possible
that, for a plausible decay rate, the OT might still be contributing
to the total light observed during the third \hst\ epoch (after 301
days, for an exposure time of 3840 seconds).  To test this, we
measured the fluxes in a small aperture on the three original images
(i.e., not subtracted) at the position of the OT; the total flux thus
contained the contribution of the host galaxy within this aperture.
We fitted an exponential decline rate plus constant to the three \hst\
fluxes (fitting a power law yields the same result)
$$
F(t) = A\exp(-t/B) + C
$$
The parameters are the half-life, normalization, and host flux within
the aperture.  It turned out that the fitted flux of the host within
the aperture (the parameter C) was smaller than the flux measured for
the third epoch. This means that indeed, the OT was contributing some
light during the third epoch.  The consequence is that when we
subtracted the third epoch from the first and second epochs, we
introduced a bias in the sense that we subtracted too much light,
since the OT was still contributing in the third epoch, used as
reference. This is almost inconsequential for the first epoch but it
is a significant fraction of the total OT flux in the second epoch.

To correct for this over-subtraction, we assumed that the decay was
exponential\footnote{The method used for this correction is as
follows. We measured fluxes on the subtracted images, $f_1$ and $f_2$;
these are the over-subtracted fluxes. The OT flux on the third image
is unknown.  The \emph{true} fluxes are denoted $F_1$, $F_2$, and
$F_3$. These fluxes satisfy the conditions $f_1 = F_1-F_3$ and $f_2 =
F_2-F_3$.  We assume that the flux decays as $F(t) \propto \exp
(-t/\tau)$.  In essence, the method is not different from a non-linear
$\chi^2$ fit: if we assume we know $F_1$ and the decay rate, we can
then compute $F_2$ and $F_3$.  From this we can compute $f_1$ and
$f_2$. If these computed values are equal to the observed values, we
have found the solution; if not, we iterate the procedure with new
values of $F_1$ and the decay rate until we reach agreement}.  The
measured and corrected magnitudes are given in Table~\ref{tbl_hst}.
We find that the decay rate is 0.0167 mag/day, corresponding to an
exponential decay time of about 60~d.  This is intermediate between
the $B$- and $V$-band decay of SN~1998bw (0.0141 and 0.0184 mag/day
respectively; McKenzie \& Schaefer 1999).  These magnitudes are in the
F606W filter which corresponds to a rest-frame wavelength of about
4900 \AA.  Since this is about half-way between the effective
wavelengths of the $B$ and $V$ filters, one can conclude that the
decay of XRF~020903 is very similar to that of SN~1998bw.

As described below (Sec.~\ref{sec_discuss}), the spectral energy
distribution (SED) evolved considerably between 6 and 38 days, but
at the time of our \hst\ observations we have no color information.
In order to facilitate the comparison of the \hst\ data with other
data, we transformed our observations in the F606W filter to the $R$
band.  We used the IRAF\footnote{IRAF is distributed by the National
Optical Astronomy Observatories, which are operated by the Association
of Universities for Research in Astronomy, Inc., under cooperative
agreement with the National Science Foundation.}/synphot task to do
this, assuming that the SED is a power law, $F_\nu \propto \nu^\beta$
(this assumption is valid given the small wavelength range we need to
consider for this procedure). We took a power-law index $\beta=-4$, as
it represents well the last SED observed (at 38 d, see
Sec.~\ref{sec_sed}).  Table~\ref{tbl_hst} gives the magnitudes we
found after correcting the F606W data to $R$ band.



\begin{table}
\begin{center}
\caption{Photometry of \hst\ data.\label{tbl_hst}}
\begin{tabular}{l c c c c c}
\tableline\tableline
UT date & Time after burst & mag\tablenotemark{a} & $\sigma$ (mag) & %
mag\tablenotemark{b} & mag\tablenotemark{c} \\
        & (days) & F606W &                        & F606W & $R$  \\
\tableline
\dataset[ADS/Sa.HST#J8IY79SMQ]{2002 Dec 03.792} &  95.426 & 24.28   & 0.05 &  24.23 & 23.81 \\
\dataset[ADS/Sa.HST#J8IYI1SFQ]{2003 Apr 23.955} & 232.824 & 27.00   & 0.12 &  26.53 & 26.11 \\
\dataset[ADS/Sa.HST#J8IYE2GJQ]{2003 Jun 30.652} & 300.625 & \nodata & 0.20 &  27.66 & 27.24 \\
\tableline
\end{tabular}

\tablenotetext{a}{Measured in the subtracted images.}

\tablenotetext{b}{These magnitudes have been corrected for
over-subtraction (see text).}

\tablenotetext{c}{$R$-band magnitudes, from F606W, assuming that the
SED is a power law ($F_\nu \propto \nu^\beta)$ of index $\beta=-4$.
The uncertainties in these magnitudes are the same as for the F606W
measurements.}

\end{center}
\end{table}

\section{Light Curve and Discussion}
\label{sec_discuss}

\subsection{The Spectral Energy Distribution\label{sec_sed}}

\begin{figure}
\epsscale{1.00}
\plotone{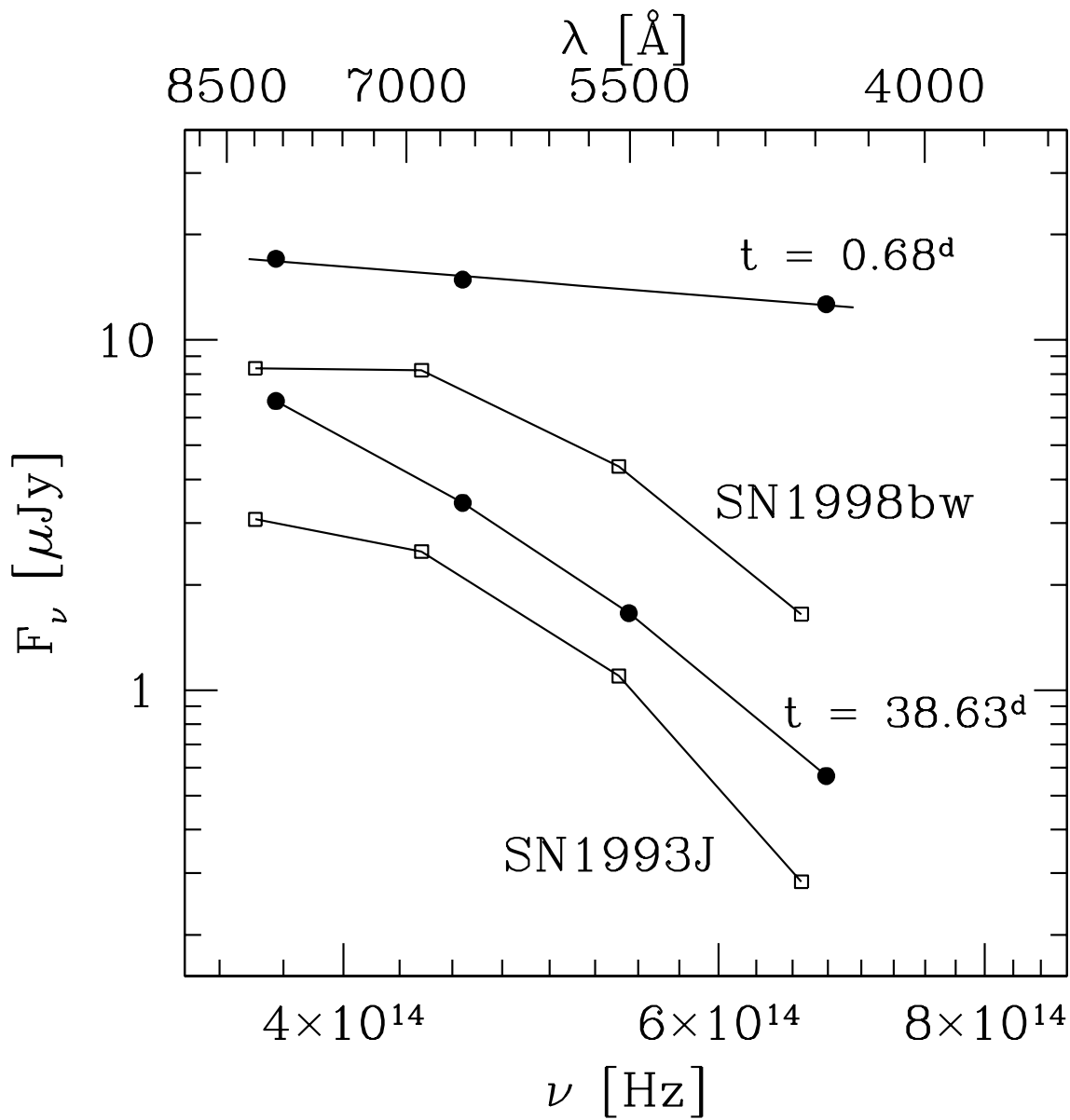}
\caption{Spectral energy distribution at 0.68~d and at 38.6~d.  The
data (black dots) have been corrected for foreground reddening. The
solid line at $t=0.68$ d is a power-law fit. The open squares are the
$UBVR$ data of two local supernovae, SN~1998bw (at 30.7~d in the rest
frame) and SN~1993J (at 30.3~d in the rest frame), appropriately
redshifted to $z=0.25$ and corrected for their respective reddening
\citep{patat2001,richmond1996a}.
\label{fig_sed}}
\end{figure}

We could determine the SED at two epochs, at 0.68~d and at 38.6~d
after the burst (see Fig.~\ref{fig_sed}).  We used magnitude zero
points from \citet{fukugita1995}. We corrected the fluxes for
foreground extinction, using a color excess $E(B-V) = 0.033$ mag from
\citet{sfd1998}.  At 0.68~d, the SED is well fitted by a power law of
index $\beta = -0.49 \pm 0.07$ (where $F_\nu \propto \nu^{\beta}$).

The striking feature of Figure~\ref{fig_sed} is the change in the
shape of the energy distribution. At 38.6~d the SED was very red.  A
formal power-law fit to the SED gives an exponent $\beta = -4.10 \pm
0.29$.  There is, however, some noticeable curvature in the SED.  In
this respect, the behavior of XRF~020903 resembles that of another
XRF, 030723 \citep{fynbo2004}, whose SED was well represented by a
power law in the first few days but after $\sim 3$ weeks it had a
strong curvature.  Figure~\ref{fig_sed} also shows that at 38.6 d the
SED resembled that of local stripped-envelope core-collapse supernovae
such as SN~1998bw and SN~1993J, although with less curvature.

\clearpage
\clearpage

\subsection{Early Light Curve\label{sec_early}}

\begin{figure}
\epsscale{1.00}
\plotone{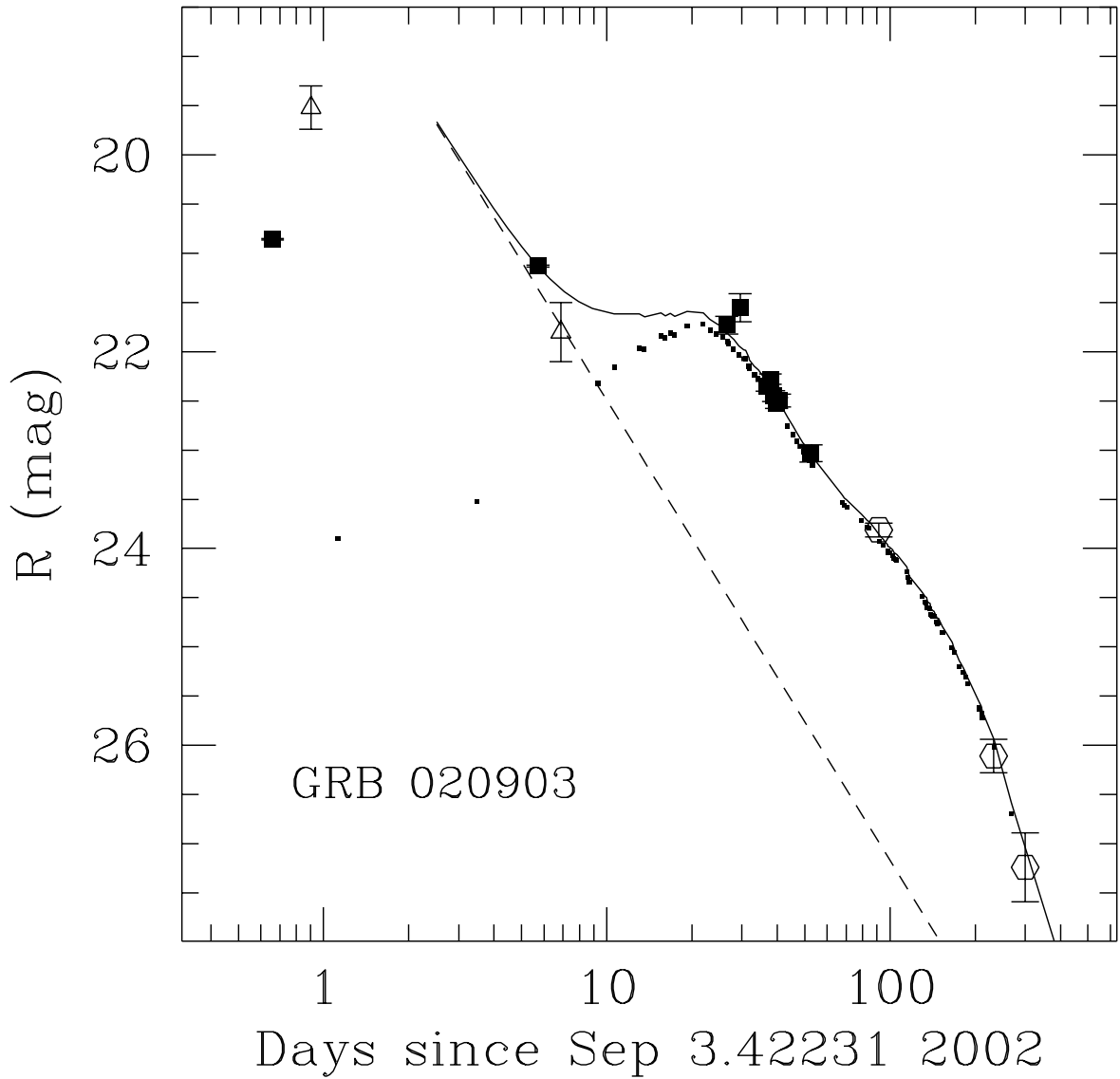}
\caption{$R$-band light curve of XRF~020903. Solid squares are our
ground-based data reported here, open triangles are the Palomar
5m measurements \citep{soderberg2004} after subtraction of the host
galaxy flux (see text for details), open hexagons (after 95 days) are
\hst\ F606W measurements transformed to the $R$ band (see text for
details on this transformation).  The small black dots are the
$V$-band light curve of SN~1998bw, corrected for reddening
\citep{patat2001}, redshifted (to account for cosmological time
dilation) then shifted by 0.8 mag (fainter).  The solid line is the
sum of a power law of index $-1.87$ (dashed line) and SN~1998bw
shifted by 0.8 mag.
\label{fig_lc}}
\end{figure}

To the data described in Sec.~\ref{sec_obs}, we added two measurements
from \citet{soderberg2004} in order to fill the large gap in the time
coverage of this burst.  We subtracted the magnitude of the host galaxy
(see Sect.~\ref{sec_host}) from their data in order to determine the
brightness of the OT.  Their last few data points are fairly noisy and
the OT contributed very little light.  The resulting light curve is
presented in Fig.~\ref{fig_lc}.

There are upper limits on the brightness of the afterglow from several
early searches \citep{gcn1531,gcn1533,gcn1535,gcn1537}.  This might be
explained when one considers the combination of the Palomar 200'' data
\citep{soderberg2004} with ours: we see that there was an initial
rise, particularly marked between our first epoch (0.68 d) and the
first observations at Palomar after 0.9 d.  Early on the afterglow was
too faint to be seen. Assuming that the rise follows a power law
($F(t) \propto t^\alpha$), the index is $+4$.

Several other bursts show this kind of behavior early on (a fast rise,
sometimes preceded by a relatively flat phase).  For instance,
GRB~970508 had a ``plateau'' after which it rose by $\sim 1.5$ mag
within $\sim 0.4$ day before decaying according to a power law
\citep{castro-tirado1998,pedersen1998,pian1998}.  It turns out that
GRB~970508 is an X-ray rich burst close to the XRF regime (this is
obtained from the integration of the spectrum given by Amati et al.
2002). GRB~030418 is also an X-ray rich GRB \citep{sakamoto2005} whose
afterglow rose during the first $\sim$ half hour
\citep{rykoff2004}. Another X-ray rich burst, GRB~041006, had a very
shallow early decay (see the data compilation in Granot et al.,
2005). The afterglow of GRB~030723, an XRF, had an almost flat light
curve during most of the first day.

All these soft bursts seem to show a trend between the softness of the
prompt emission and the early decay rate of the afterglow.  From a
theoretical point of view, \citet{granot2002} and \citet{gk2003} show
that, depending on the viewing angle it is possible to have a rapidly
rising afterglow, such as is observed for XRF~020903.  In the models
of \citet{gk2003} and \citet{granot2005} the shape of the light curve
is determined in particular by the opening angle of the jet, the
viewing angle (or rather, the relation between these two angles), and
the angular distribution of energy across the jet.  In their model,
X-ray flashes are seen off-axis, i.e., the viewing angle is larger than
the opening angle of the jet.  Using this approach, \citet{granot2005}
can reproduce the light curves of XRF~030723 and GRB~041006 fairly
well.  For XRF~020903, we can say in particular that in order to have
a fast rise, one needs a jet with a very sharp edge
\citep{gk2003,granot2005}.

Because of the large gap between 0.9 d and 5.7 d after the burst, it
is impossible to know when the OT reached its maximum brightness.  It
is equally impossible to know its decay rate.  If we assume that the
OT reached maximum brightness at 0.9 d and also assume that it
subsequently decayed following a power-law ($f(t) \propto t^\alpha$),
then the index is $\alpha = -0.8$ in this 5 day time interval.  If the
afterglow continued to rise after 0.9 d then the later decay will be
faster than $-0.8$. Given the paucity of data in the first week, we
cannot rule a more complex behavior of the afterglow. In particular,
the second Palomar 5m data point at 6.9 d implies a faster decay
(although this measurement has a large uncertainty).

For some late-time images (obtained later than 25 days after the
burst), particularly in the $B$ and $V$ passbands, the optical
transient was not detected in the subtracted frames, we could only set
upper limits to the magnitudes (see Table~\ref{tbl_phot}). From $R$
band data one can nevertheless say that the OT must have decayed very
slowly between 6 and 26 d. After that time it became fainter at a fast
rate. This decay continued at least until 300 d after the burst, as
evidenced by the \hst\ data (see Fig.~\ref{fig_lc}).

\clearpage
\clearpage

\subsection{Late-Time Light Curve\label{sec_late}}

The OT decayed by $\sim 0.5$ mag between 5 d and 26 d, after which the
flux decreased at a faster rate.  Figure~\ref{fig_lc} shows
that a simple broken power-law model will fail to describe the light
curve; the addition of a supernova component is necessary to fit the
late-time bump.  For the fit, we corrected the magnitudes for the
foreground reddening. The SN~1998bw data were also corrected for the
reddening using the value of \citet{patat2001}.

We tried to reproduce the light curve by fitting a combination of
power law (with two free parameters - decay index and flux
normalisation), and shifted SN~1998bw light curve (adding two more
free parameters - magnitude shift and time stretch factor; see
e.g. Zeh et al., 2004).  Concentrating on the data obtained after 5
days, the fit yields the following results: the power law has an index
equal to $\alpha=-1.87 \pm 0.28$; the supernova light curve is that of
SN~1998bw in the $V$ band\footnote{We used the $V$-band light curve of
SN~1998bw because at a redshift of 0.251, the observed $R$ band is
almost exactly equal to the rest-frame $V$ band.}  shifted by $0.8 \pm
0.08$ mag (fainter).  A stretch factor was also applied to the SN
light curve although it does not appear necessary ($1.0 \pm 0.03$).

We tried other SN light curves, in particular those of SN Ib 1993J
\citep{richmond1996a} and SN Ic 1994I \citep{richmond1996b}.  Both
gave substantially poorer fits than SN~1998bw. We conclude that the
supernova associated with XRF~020903 was very similar to SN~1998bw,
except that it was fainter and redder (see Fig.~\ref{fig_sed}). The SN
could actually be brighter than the limit derived above if there is
some internal reddening.  These results are consistent with, but more
precise than, those of \citet{soderberg2005a}.  They found that the
supernova is $0.6 \pm 0.5$ mag fainter than SN~1998bw, and that the SN
light curve had to be ``compressed'', i.e., a faster decay than
SN~1998bw.  We believe that our better photometry (obtained via image
subtraction, as opposed to simply extracted from the GRB Coordinates
Network Circulars), and the correction of \hst\ data for
over-subtraction explain these differences.

Several supernovae associated with GRBs are sufficiently well
characterized (see, e.g., Zeh et al. 2004) that one can plot the SN
luminosity (in units of SN~1998bw) versus the SN decay time.
\citet{stanek2005} hinted at a correlation between these two
quantities.  With a luminosity ratio of $\sim 0.5$ and a stretch
factor of 1, the SN associated with XRF~020903 adds to the dispersion
in this plot, and does not follow the trend seen for Type Ia SNe
\citep{phillips1993}.  A large amount of reddening ($E(B-V) \sim
0.25$) would bring this SN in line with the trend but there is no
evidence for this amount of reddening.  Actually, with $E(B-V)=0.165$,
the reddening-corrected SED at 0.68 d would be flat (slope=0).  We
consequently take this value of $E(B-V)$ as a strong upper limit.
Were the reddening so large (0.165), then the SN brightness at maximum
would only be 0.3 mag fainter than SN~1998bw. As noted above however,
this is very strong upper limit and this transform into a strong upper
limit on the peak magnitude of the SN associated with XRF~020903.

\subsection{The Host Galaxy\label{sec_host}}

$BVRI$ host magnitudes (see Table~\ref{tbl_host}) were obtained from
late-time ground-based images.  Even though these images were obtained
in good seeing conditions, one should be aware that the host is at
best marginally resolved; we can not distinguish the various knots
seen in the right panel of Fig.~\ref{fig_host}.  Our host magnitudes
are in good agreement with values reported in the Gamma ray bursts
Coordinates Network (GCN) after accounting for the non-negligible
fraction of the light coming from the OT in the GCN magnitudes
\citep{gcn1563,gcn1631}.  In particular, our $R$-band magnitude is
very close to that given by \citet{soderberg2004}.  The uncertainties
in the host magnitudes are large because of the presence of the
complex of galaxies ``G2+G3'' which contaminates the photometry in
ground-based data.  The fact that photometry is delicate can also be
inferred from the dispersion in the magnitudes in \citet{gcn1563} and
\citet{gcn1631}.  We conservatively estimate our host photometry
uncertainties to be $\pm 0.1$ mag.


\begin{table}
\begin{center}
\caption{Photometry of the host galaxy.\label{tbl_host}}
\begin{tabular}{l l c l c}
\tableline\tableline
UT date & Telescope\tablenotemark{a} & Filter & mag & $\sigma$ (mag) \\
\tableline
2004 Aug 24.259 & DK 1.5m & B & 21.7 & 0.1 \\
2004 Aug 22.278 & DK 1.5m & V & 20.8 & 0.1 \\
2004 Sep 14.032 & CTIO 4m & R & 20.8 & 0.1 \\
2004 Sep 14.059 & CTIO 4m & I & 20.5 & 0.1 \\
\tableline
\end{tabular}

\tablenotetext{a}{DK 1.5 = 1.5~m Danish telescope at La Silla,
CTIO 4m = Cerro Tololo 4~m}

\end{center}
\end{table}

\subsection{Spectroscopy\label{sec_spec}}

A spectrum was obtained on 8.4 October 2002 (35~d after the burst)
using the Keck-I telescope and the Low Resolution Imaging Spectrometer
(LRIS; Oke et al. 1995).  We used the 400/3400 grism on the blue side;
on the red side we used the 400/8400 grating.  The slit width was
1\arcsec, aligned at a position angle of 162.6$\arcdeg$, which was the
parallactic angle \citep{filippenko1982} at the time of the
observations.  We obtained two exposures, one of 1250~s and the other
of 1800~s. The conditions were poor, making the absolute flux
calibration quite uncertain, although relative fluxes should be little
affected.

\clearpage
\begin{figure}
\epsscale{1.00}
\plotone{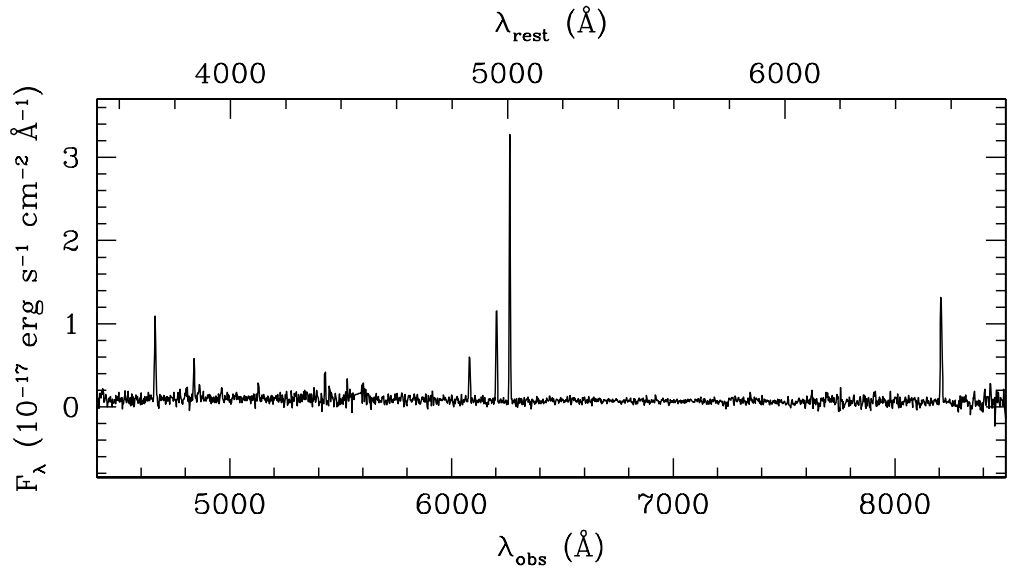}
\caption{Keck spectrum of the host of XRF~020903.  We detect strong
emission lines associated with star formation in a low-metallicity
environment. The fluxes have been corrected for foreground extinction.
\label{fig_spec}}
\end{figure}
\clearpage

We clearly detect several lines associated with strong, on-going star
formation, [O~II] $\lambda$3727, [Ne~III] $\lambda 3867$, hydrogen
Balmer lines, and [O~III] $\lambda\lambda$4959, 5007.  Several common
lines are not detected, in particular [N~II] $\lambda\lambda$6548,
6583.  Before measuring fluxes, we corrected the spectrum for the
small foreground reddening, $E(B-V) = 0.033$ mag \citep{sfd1998}.
Fluxes and equivalent widths were then measured by fitting a Gaussian
to the line profiles, and the results are given in
Table~\ref{tbl_spec}.  Using eight strong lines, we measured a
redshift $z = 0.2506 \pm 0.0003$, in agreement with the value
determined by \citet{soderberg2004}.

\clearpage 
\begin{table}
\begin{center}
\caption{Emission line fluxes.\label{tbl_spec}}
\begin{tabular}{l c c c c}
\tableline\tableline
Ion & Wavelength                & Flux    &     & Equivalent width \\
    & $\lambda_{\rm obs}$ (\AA) & $f$ & $f/f_{H\beta}$ & EW$_{\rm rest}$ (\AA) \\
\tableline
$[$O~II$]$   & 4663.52 &   7.982  & 2.274 &  $-68.96$ \\
$[$Ne~III$]$ & 4838.76 &   2.923  & 0.833 &  $-24.47$ \\
$[$Ne~III$]$ & 4963.32 &   1.059  & 0.302 &   $-8.67$ \\
H$\delta$    & 5129.45 &   1.272  & 0.362 &  $-10.12$ \\
H$\gamma$    & 5430.13 &   2.019  & 0.575 &  $-15.94$ \\
H$\beta$     & 6081.46 &   3.510  & 1.000 &  $-36.08$ \\
$[$O~III$]$  & 6203.46 &   7.386  & 2.104 &  $-77.44$ \\
$[$O~III$]$  & 6263.21 &   20.89  & 5.952 & $-221.16$ \\
H$\alpha$    & 8208.78 &   11.46  & 3.265 & $-152.40$ \\
$[$N~II$]$   & 8239.60 & $<0.133$ &   --  &    --   \\
\tableline
\end{tabular}
 
\tablecomments{Fluxes are in units of $10^{-17}$
erg~s$^{-1}$~cm$^{-2}$. They have been corrected for foreground
reddening (using $E(B-V)=0.033$, see Sect.~\ref{sec_sed}).  Equivalent
widths have been corrected to the rest frame.}
 
\end{center}
\end{table}
\clearpage

We set out to determine the metallicity of the host galaxy from the
emission-line ratio \R23, following the procedure outlined in
\citet{kd2002} as updated in \citet{kk2004}. This ratio is defined as
$R_{23} \equiv (I_{\mbox{[O {\sc II}]}\lambda 3727} + 
I_{\mbox{[O {\sc III}]}\lambda 4959} + 
I_{\mbox{[O {\sc III}]}\lambda 5007})/I_{\mathrm H\beta}$.
Using the values in Table~\ref{tbl_spec}, we find $\log R_{23} =
1.014$. This is beyond the range of values for which the \R23 method
has been calibrated (see, e.g., Kewley \& Dopita 2002; Kobulnicky \&
Kewley 2004).  Consequently, we cannot use the \R23 ratio to determine
an accurate abundance.

There are several possible reasons explaining why the \R23 ratio is
high.  One is that the average density in the host of XRF~020903 is
higher than in the models of \citet{kd2002}. This is sometimes
observed for Lyman-break galaxies (L. Kewley, 2005, private communication).
Another reason is that the excitation process may not be purely due to
photoionization. In particular, shock excitation would change the line
ratios, seriously compromising the \R23 method.  We can nevertheless
obtain an upper limit on the abundance [O/H], using the fact that the
[N~II] $\lambda 6583$ line is not detected. We estimated the average
noise ($\sigma$) in the wavelength region around this line, and took
the upper limit on the line flux as $3\sigma = 1.33\times 10^{-18}$
erg~s$^{-1}$~cm$^{-2}$. From the ratio of $\log$([N~II]/H$\alpha) \leq
-2.02$, we obtain an upper limit on the abundance, namely [O/H]$\leq
-0.6$ dex (see Figure~7 of Kewley \& Dopita 2002).

\section{Conclusions\label{sec_concl}}

The early afterglow of XRF~020903 behaved rather unexpectedly, as it
rose in the first day. This type of time evolution has already been
seen in a few other GRB/XRF afterglows.  Some models can actually
reproduce this behavior. In particular, \citet{granot2005} attribute
this to the fact that the burst is seen away from the symmetry axis of
the jet (Sec.~\ref{sec_early}). A rising light curve is a prediction
of the off-axis model and it is interesting to note that, as the most
extreme object of this class, XRF~020903 also has a very fast early
rise.  The afterglow light curves of several other soft bursts showing
a plateau or a rising phase may be explained in the framework of this
model.  The behavior of these bursts shows that there may then be a
relation between the softness of the burst and the shape of the early
optical light curve of the afterglow.

The optical transient was still fairly bright after $\sim 30$~d. The
late-time evolution of XRF~020903 is well described by the light curve
of SN~1998bw, shifted by 0.8 mag (fainter).  The other striking
feature is the very red SED at 38~d after the burst ($\sim 31$~d in
the rest frame). A supernova seems to be the only way to have such a
red and curved SED, as other explanations predict a power law for the
SED.  Assuming for a moment that there is no SN and that the late-time
fading is a power law, it is very surprising that a local supernova,
SN~1998bw, would fit the light curve \emph{and} the SED so
well. Furthermore, it is a challenge to GRB models to explain how the
afterglow could have remained so bright for over 30 days (decaying
only by $\sim 0.5$~mag in 20 days) and be so red.  In summary, the
existence and brightness of the bump, its timing, and the SED at 38~d,
can all be explained naturally by a SN. Adding this to the fact that a
spectrum of the OT at 24.6~d strongly resembles a spectrum of
SN~1998bw \citep{soderberg2005a}, there is no doubt that XRF~020903 is
associated with a core-collapse supernova.

The global properties of the host galaxy are similar to those of other
GRB host galaxies, in the sense that it is subluminous, actively
star-forming, and fairly metal-poor.  It resembles the host of
GRB~031203 (Prochaska et al., 2004; see also Sollerman et al., 2005).

It is unfortunate that the afterglow was not detected at early times.
After re-analysis of early data, the announcement of an afterglow at
optical and radio wavelengths \citep{soderberg2004} came at a time
when the host was already dominating the optical flux. This made
photometry difficult and most groups stopped taking data.  It is
conceivable that the afterglow may have been missed at early times
because of the presence of the host.  Even in the presence of a bright
host, it is worth taking data for a long time after the burst, since
image subtraction can provide reliable light curves.  The X-ray
telescope (XRT) on Swift provides positions good enough that the
emergence of a supernova in the XRT error box after $\sim 2$ weeks
would pinpoint the location of the host galaxy.  GRB~020410 is a
perfect example of this strategy \citep{levan2005b}.

While there is clearly a SN associated with XRF~020903, there are
several XRFs where the evidence for a SN is weak, if not lacking
altogether.  \citet{levan2005a} and \citet{soderberg2005a} constrain
the brightness of any SN associated with several X-ray flashes,
although the lack of measured redshifts for some objects make these
constraints less strong than one would wish. XRF~030723 had a strong
bump on its light curve \citep{fynbo2004} that is interpreted as a SN,
akin to SN~1994I, and a SN similar to SN~1998bw can be confidently
excluded for that XRF.  All this shows that if every XRF has
an underlying SN, they have a broad range of peak luminosities
\citep{soderberg2005a} and light curve shapes.
This being said, there is now at least one good example of a classical
GRB (e.g. Stanek et al. 2003, Hjorth et al.  2003, Matheson et al.
2003 for GRB~030329), X-ray rich GRB (Stanek et al.  2005, Soderberg
et al 2005, for GRB~041006), and X-ray flash (this paper and Soderberg
et al., 2005, for XRF~020903) clearly associated with supernovae.
This is predicted by theoretical models of stellar collapse
\citep{zwh2004} and by GRB models where the difference between these
various events is explained by a different viewing geometry.  All this
reinforces the suspicion that XRFs, X-ray rich GRBs, and long GRBs are
slightly different outcomes of the same phenomenon: the collapse of a
massive star. The physical mechanisms powering these events appear to
be very similar, albeit with a fairly wide variety of observational
properties.

\acknowledgments

We thank the referee for constructive comments that helped us improve
the paper.   We thank L. Kewley and J. Granot for very helpful
discussions.
Support for program GO-9405 was provided by NASA through a grant from
the Space Telescope Science Institute, which is operated by the
Association of Universities for Research in Astronomy, Inc., under
NASA contract NAS 5-26555.
This paper is partly based on observations made with the Italian
Telescopio Nazionale Galileo (TNG) operated on the island of La Palma
by the Fundaci{\'o}n Galileo Galilei of the INAF (Instituto Nazionale
di Astrofisica) at the Spanish Observatorio del Roque de los Muchachos
of the Instituto de Astrof\'{\i}sica de Canarias.
We also acknowledge the valuable efforts of the Cerro Tololo
Inter-american Observatory, Asiago and TNG personnel.
The research of J. Gorosabel is partially supported by the Spanish
Ministry of Science and Education through programs
ESP2002-04124-C03-01 and AYA2004-01515 (including FEDER funds).
The work of A.V.F. is supported by NSF grant AST-0307894; he is also
grateful for a Miller Research Professorship at U.C. Berkeley, during
which part of this work was completed.




\end{document}